\documentclass[reprint,amsmath,amssymb,aps,superscriptaddress,showpacs,floatfix,byrevtex,longbibliography]{revtex4-1}

\usepackage{amsmath}
\usepackage{amssymb}
\usepackage{amstext}
\usepackage{amsopn}
\usepackage{amsfonts}
\usepackage{amsxtra}
\usepackage[english]{babel}
\usepackage{graphicx}
\usepackage{float}
\usepackage{bm}
\usepackage{multirow}
\usepackage{dcolumn}
\usepackage{color}
\usepackage{hyperref}

\begin{document}

\preprint{Draft --- not for distribution}

%
%
\title{Unusual electronic and vibrational properties in the \\
  colossal thermopower material FeSb$_\mathbf{2}$}
\author{C. C. Homes}
\email{homes@bnl.gov}
\affiliation{Condensed Matter Physics and Materials Science Division,
  Brookhaven National Laboratory, Upton, New York 11973, USA}
\author{Q. Du}
\affiliation{Condensed Matter Physics and Materials Science Division,
  Brookhaven National Laboratory, Upton, New York 11973, USA}
\affiliation{Department of Materials Science and Chemical Engineering,
  Stony Brook University, Stony Brook, New York 11790, USA}
\author{C. Petrovic}
\affiliation{Condensed Matter Physics and Materials Science Division,
  Brookhaven National Laboratory, Upton, New York 11973, USA}
\affiliation{Department of Materials Science and Chemical Engineering,
  Stony Brook University, Stony Brook, New York 11790, USA}
\author{W. H. Brito}
\author{S. Choi}
\affiliation{Condensed Matter Physics and Materials Science Division,
  Brookhaven National Laboratory, Upton, New York 11973, USA}
\author{G. Kotliar}
\affiliation{Department of Physics and Astronomy, Rutgers,
  The State University of New Jersey, Piscataway, New Jersey 08854, USA}

%
%
%
%
\begin{abstract}
The iron antimonide FeSb$_2$ possesses an extraordinarily high thermoelectric power
factor at low temperature, making it a leading candidate for cryogenic thermoelectric
cooling devices.  However, the origin of this unusual behavior is controversial, having
been variously attributed to electronic correlations as well as the phonon-drag
effect.  The optical properties of a material provide information on both the electronic
and vibrational properties.  The optical conductivity reveals an anisotropic response at
room temperature; the low-frequency optical conductivity decreases rapidly with temperature,
signalling a metal-insulator transition.  One-dimensional semiconducting behavior
is observed along the $b$ axis at low temperature, in agreement with first-principle
calculations.
The infrared-active lattice vibrations are also symmetric and extremely narrow, indicating
long phonon relaxation times and a lack of electron-phonon coupling.  Surprisingly,
there are more lattice modes along the $a$ axis than are predicted from group theory;
several of these modes undergo significant changes below about 100~K, hinting at a
weak structural distortion or phase transition.
While the extremely narrow phonon line shapes favor the phonon-drag effect, the
one-dimensional behavior of this system at low temperature may also contribute
to the extraordinarily high thermopower observed in this material.
\end{abstract}
%
%
\maketitle

{\em Introduction.} FeSb$_2$ crystallizes into an orthorhombic structure with two formula units per
unit cell, as shown in Fig.~1(a).  Despite this simple structure, there are two moieties
of FeSb$_2$ crystals, those with a putative metal-insulator transition (MIT) in which the dc
conductivity along the \emph{b} axis first increases below room temperature, reaching a broad
maximum at about $80 - 100$~K, before decreasing dramatically as the temperature is further
reduced\cite{jie12}, and a second class of materials without a MIT in which the dc conductivity
immediately begins to decrease as the temperature is lowered \cite{bentien07,sun09,sun10,jie12},
as shown in Fig.~1(b), for the two types of crystals examined in this work.  Both classes of
materials have a high thermoelectric power factor at low temperature; however, it is
extraordinarily high in the materials with a MIT\cite{jie12}.
The thermoelectric efficiency is given by the dimensionless figure of merit $ZT=\sigma S^2T/\kappa$,
where $\sigma$, $S$, $T$, and $\kappa$ are the conductivity, Seebeck coefficient, temperature, and
thermal conductivity, respectively; the thermoelectric power is simply $S^2\sigma$; in FeSb$_2$ the
Seebeck coefficient may be as high as $S\simeq -45$~mV$\,$K$^{-1}$ at low temperature, resulting in
the highest power factor ever recorded \cite{bentien07}.  In general, there are two strategies
for increasing $ZT$; reduce $\kappa$ or increase the power factor $S^2\sigma$.  However, because
the source of this large thermoelectric response is not entirely understood, with electronic
correlations \cite{petrovic05,perucchi06,bentien07,sun09,sun10,herzog10,jie12,figueira12,
fuccillo13,sun13,takahashi16}, as well as the phonon-drag effect \cite{tomczak10,pokharel13,
liao14,battiato15,takahashi16}, having been proposed, it is not clear which approach offers
the best chance of success.
%
%
%
\begin{figure}[t]
\includegraphics[width=3.20in]{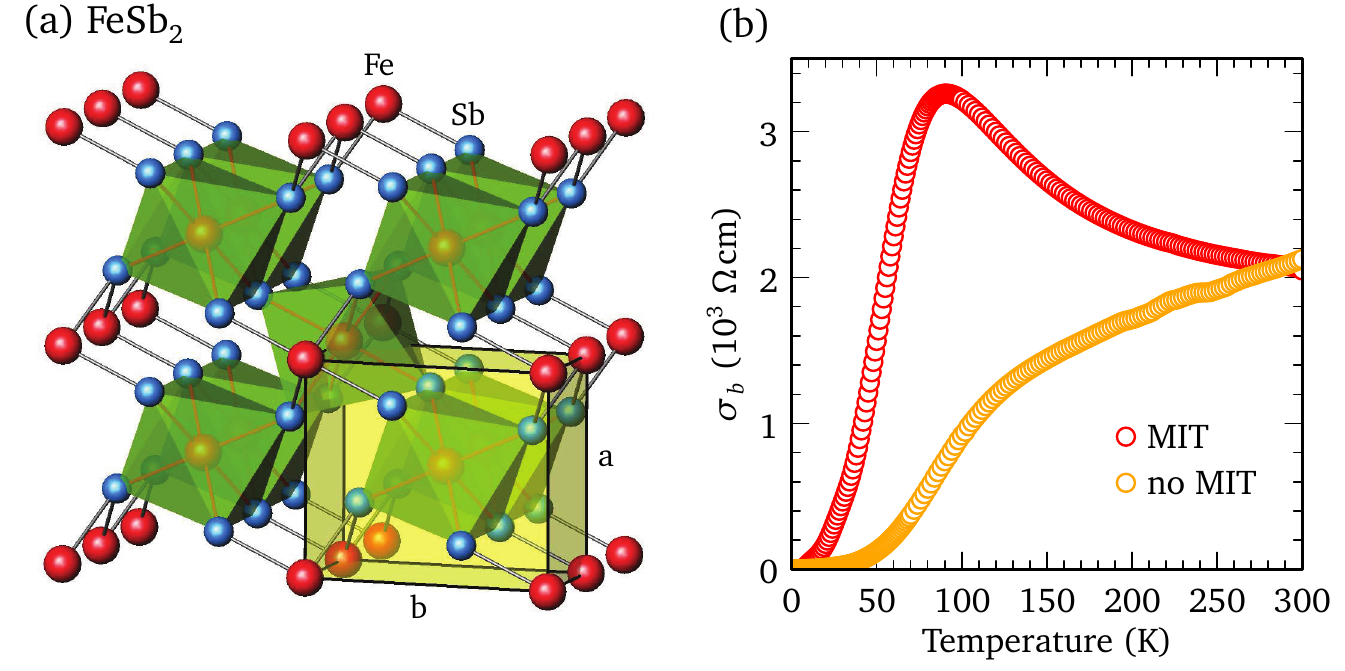}
\caption{ Structural and transport properties of FeSb$_{2}$.
(a) The crystal structure of FeSb$_2$ in the orthorhombic $Pnnm$ (58) setting
is shown for an \emph{a--b} face, with the $c$ axis facing into the paper; there are
two formula units per unit cell.  The orthorhombic unit cell dimensions are roughly
5.83, 6.53 and 3.20~\AA\ for the $a$, $b$, and $c$ axis, respectively\cite{holseth68}.
The fractional coordinates are Fe $(0,0,0)$ and Sb $(x,y,0)$, with $x=0.1885$ and $y=0.3561$.
Each Fe atom sits at the center of a deformed octahedra which share edges along the $c$ axis.
(b) The dc conductivity along the \emph{b} axis, determined from the dc resistivity
$\sigma_b = 1/\rho_b$, is shown for the crystal that displays a MIT, and one that does not.
The dc transport properties have been measured on the same samples that the optical
measurements were performed. }
%
\label{fig:ca122}
\end{figure}

%
%
\begin{figure}[thb]
\includegraphics[width=2.49in]{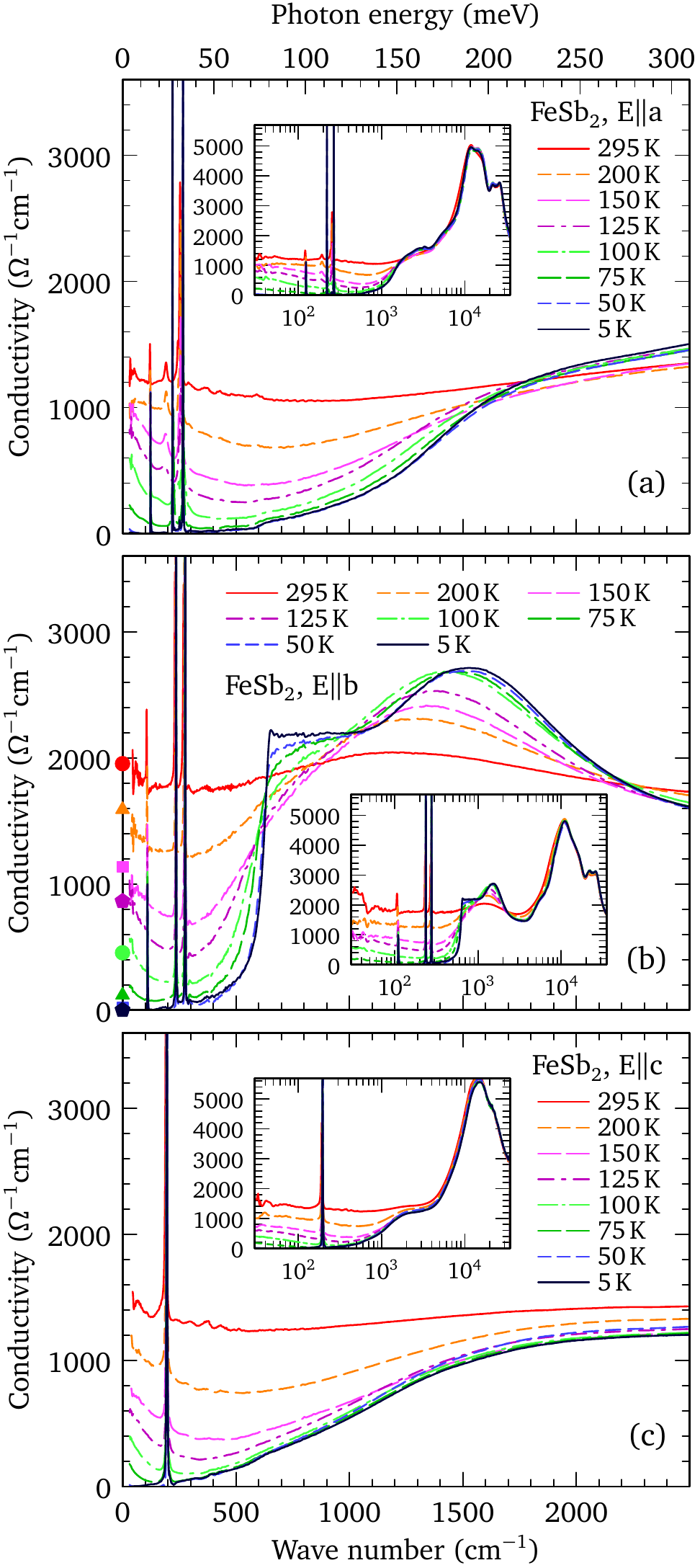}
\caption{The optical conductivity of FeSb$_{2}$.
(a) The temperature dependence of the real part of the optical
conductivity for light polarized along the \emph{a} axis.
Inset: the conductivity shown over a wide energy range.
(b) The temperature dependence of the optical conductivity for light
polarized along the \emph{b} axis.  As the temperature is reduced the
low-frequency conductivity decreases dramatically revealing a step-like
feature at $\simeq 600$~cm$^{-1}$; three narrow infrared-active lattice modes
all lie below this energy.    The points on the conductivity axis correspond
the values for $\sigma_{dc}$ measured along this direction in a sample without
a MIT and normalized to the extrapolated value for $\sigma_1(\omega\rightarrow 0)$
at room temperature.
Inset: the conductivity shown over a much larger energy range.
(c) The temperature dependence of the optical conductivity along
the \emph{c} axis, which is similar in magnitude to the conductivity along
the \emph{a} axis.
Inset: the conductivity shown over a wide energy range.}
\vspace*{-0.5cm}%
\end{figure}

%
The complex optical properties yield information about both the electronic and vibrational
properties of a material, and can offer insights into the origin this unusual behavior.
The real part of the optical conductivity is particularly useful as it yields information
about the gapping of the spectrum of excitations in systems with a MIT, and in the
zero-frequency limit, the dc conductivity is recovered, $\sigma_1(\omega\rightarrow 0)
\equiv \sigma_{\rm dc}$, allowing comparisons to be made with transport data.  Furthermore,
the infrared-active transverse-optic modes at the center of the Brillouin zone may be
observed in $\sigma_1(\omega)$ as resonances superimposed upon an electronic background
(or antiresonances if strong electron-phonon coupling is present).  The optical properties
of FeSb$_2$ have been previously examined in the \emph{a-b} planes \cite{perucchi06} and
along the \emph{c} axis \cite{herzog10}, revealing a semiconducting response at low
temperature and evidence for electron-phonon coupling.

%
%
%
{\em Results.} Crystals of FeSb$_2$ have been prepared by the usual methods\cite{petrovic03,bentien06}.
The reflectance of several single crystals, with and without a MIT, has been measured over
a wide frequency range ($\simeq 3$~meV to 4~eV) at a variety of temperatures for light polarized
along the \emph{a}, \emph{b}, and \emph{c} axes \cite{homes93} (Supplementary Fig.~S1).
Only naturally-occurring crystal faces have been examined, although after an initial measurement
the \emph{c} axis face was polished to remove some surface irregularities.  Polishing broadens
the lattice mode(s), but does not otherwise affect the optical properties.  After the optical
measurements were completed, the samples were dismounted and the dc resistivity, $\rho_{dc}$,
was measured using a standard four-probe technique \cite{jie12} [the dc conductivity, $\sigma_{dc}
= 1/\rho_{dc}$, is shown along the $b$ axis in Fig.~1(b)].
%
%

\noindent While the reflectance is a tremendously useful quantity, it is a combination of
the real and imaginary parts of the dielectric function, and as such it is not necessarily
intuitive or easily understood.
It is much simpler to examine the real part of the optical conductivity, determined from a
Kramers-Kronig analysis of the reflectance, \cite{dressel-book} shown in the infrared region
along the \emph{a}, \emph{b}, and \emph{c} axes Figs.~2(a), (b), and (c) , respectively;
the insets show the conductivity over a much wider frequency range.   Interestingly, the
temperature dependence of the reflectance for crystals with and without an MIT is identical
in the infrared region (shown for light polarized along the \emph{b} axis in Supplementary
Fig.~S2).  Consequently, the low-frequency optical conductivity in Fig.~2 never shows the
initial increase with decreasing temperature that is seen in the dc conductivity in
samples with a MIT; instead, the low-frequency optical conductivity decreases with
temperature along all three lattice directions, suggesting that no MIT is present.
The apparent dichotomy between the temperature dependence of the dc resistivity and the
optical conductivity in crystals with an MIT [Figs.~1(b) and S2(a)] indicates that the
dc transport properties are being driven by an impurity band that is sufficiently narrow
so that its response falls below our lowest measured frequency.

%
%
\noindent At room temperature, the real part of the optical conductivity may be described
by a simple Drude model with Fano-shaped Lorentz oscillators to describe possible
electron-phonon coupling,\cite{homes16}
\begin{widetext}
\begin{equation}
  \sigma_1(\omega) = \frac{2\pi}{Z_0} \left[ \frac{\omega_p^2\,\tau}{(1+\omega^2\tau^2)} +
  \sum_j \frac{\Omega_j^2\left[ \gamma_j\omega^2 - 2(\omega^2\omega_j-\omega_j^3)/q_j -
    \gamma_j\omega_j^2/q_j^2\right]}  {(\omega^2 - \omega_j^2)^2+\gamma_j^2 \omega^2} \right],
\end{equation}
\end{widetext}
where the first term denotes the (Drude) free carriers, with the square of the plasma
frequency $\omega_p^2 = 4\pi ne^2/m^\ast$ and scattering rate $1/\tau$, where $n$ and
$m^\ast$ are the carrier concentration and effective mass, respectively.  The second
term is a summation of oscillators with position $\omega_j$, width $\gamma_j$, strength
$\Omega_j$, and (dimensionless) asymmetry parameter $1/q_j^2$, that describe the vibrations
of the lattice or bound excitations (interband transitions); $Z_0\simeq 377$~$\Omega$ is
the impedance of free space, yielding units for the conductivity of $\Omega^{-1}$cm$^{-1}$.
In the $1/q^2\rightarrow 0$ limit a symmetric Lorentzian profile is recovered; however,
as $1/q^2$ increases the line shape becomes increasingly asymmetric.
The real part of the optical conductivity along the \emph{a} and \emph{c} axes at 295~K,
shown in Figs.~2(a) and (c), respectively, are similar, with $\sigma_{dc}\equiv \sigma_1(\omega\rightarrow 0)
\simeq 1400$~$\Omega^{-1}{\rm cm}^{-1}$.  Along the \emph{b} axis the optical conductivity
at room temperature is higher, with $\sigma_{dc}\simeq 1900$~$\Omega^{-1}{\rm cm}^{-1}$
[Fig.~2(b)].
Fits to the optical conductivity at 295~K describe the data quite well and yield
$\omega_{p}\simeq 7690$, 6770 and 6480~cm$^{-1}$, and $1/\tau \simeq 800$, 400 and
490~cm$^{-1}$ along the \emph{a}, \emph{b}, and \emph{c} axes, respectively
(Supplementary Fig.~S3); the optical conductivity along the
\emph{b} axis at room temperature is shown in Fig.~3(a).  This anisotropy suggests that
$m^\ast$ is slightly lower along the \emph{a} axis, and that the larger value for
$\sigma_{dc}$ along the \emph{b} axis is a consequence of a lower scattering rate
(Supplementary Table 1).
As Fig.~2 indicates, the Drude component begins to decrease rapidly in strength below
room temperature, along all three directions, with a commensurate loss of spectral
weight (the area under the conductivity curve) that is transferred from low to high
frequency \cite{perucchi06}.  The Drude  model may be used to track the temperature
dependence of $\omega_p$ and $1/\tau$ down to about 75~K, below which the free-carrier
response becomes too small to observe in our measurements.  The Drude expression for
the dc conductivity, $\sigma_{dc} = 2\pi\omega_p^2\tau/Z_0$, decreases rapidly as the
temperature is lowered, suggesting that the transport may be described by an activation
energy $E_a$ using the Arrhenius equation,
\begin{equation}
   \sigma_{dc} \propto \omega_p^2\tau =A\,e^{-E_a/(k_{\rm B}T)},
\end{equation}
where $E_a=E_g/2$.  Transport measurements typically identify two gaps in FeSb$_2$,
$E_g\simeq 5$~meV below about 20~K, and $E_g\simeq 26 - 36$~meV in the $50 - 100$~K
temperature range \cite{bentien07,sun09,sun10,jie12}.  The Arrhenius relation describes
the temperature dependence of $\sigma_{dc}$ along all three lattice directions quite
well (see Supplementary Fig.~S4), and yields values for the transport gap of
$E_g \simeq 20.6$, 19.5 and $24.8\pm2$~meV along the \emph{a}, \emph{b},
and \emph{c} axes, respectively, in good agreement with the high-temperature values
for the transport gap.

%
%
\begin{figure}
\includegraphics[width=2.60in]{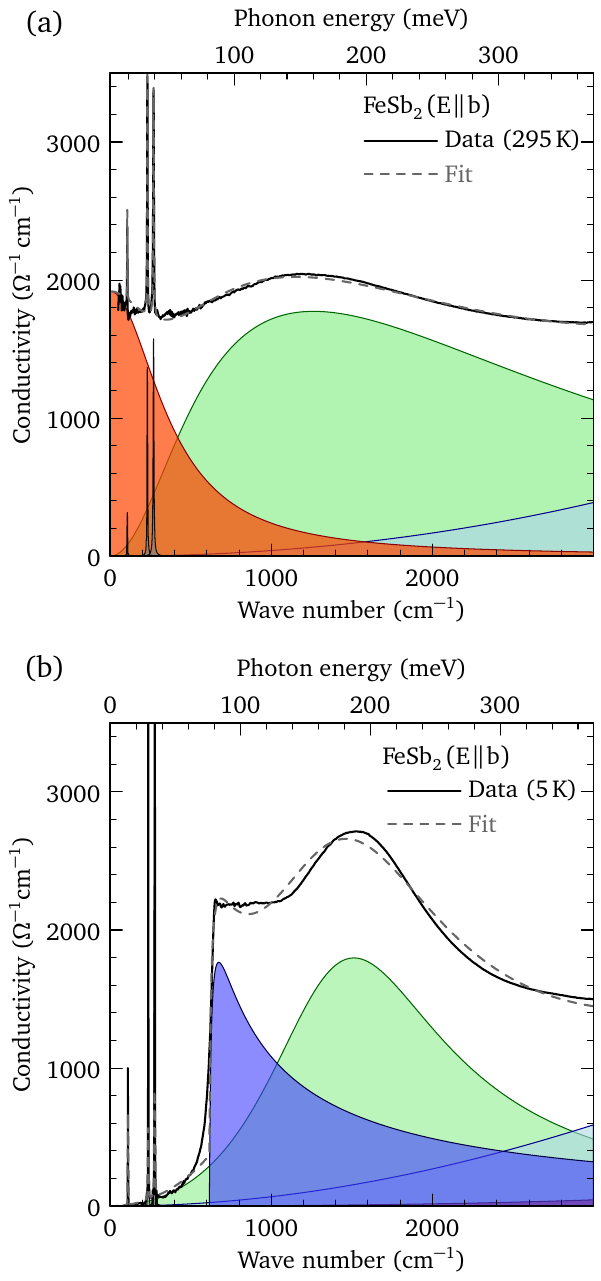}
\caption{Singular behavior in FeSb$_{2}$.
(a) The fit to the real part of the optical conductivity of FeSb$_2$ along
the \emph{b} axis at 295~K using a Drude (free carrier) component; Lorentzian
oscillators with Fano profiles have been used to describe the lattice modes,
while the interband (bound) excitations are assumed to be symmetric.
(b) The fit to the gapped optical conductivity of FeSb$_2$ along the
\emph{b} axis at 5~K using Fano-shaped Lorentzian oscillators to describe
the lattice modes, and symmetric profiles for the interband (bound) excitations,
in linear combination with the optical conductivity expected for a one-dimensional
semiconductor with a characteristic $1/\sqrt{w}$ singularity above the semiconducting
optical gap $2\Delta$.}
\end{figure}

%
%
%
{\em Discussion.} While the Drude model with Fano-shaped Lorentz oscillators is able to reproduce the
temperature-dependence of the optical conductivity along the \emph{a} and \emph{c} axes
reasonably well, it fails to describe the sharp feature that develops along the \emph{b}
axis at low temperature.  This step-like feature is the signature of a van Hove singularity
in the density of states.  The asymmetric profile in the real part of the low-temperature
optical conductivity resembles the $1/\sqrt{\omega}$ singularity response observed in
one-dimensional semiconductors,
\begin{equation}
  \sigma_{1{\rm D}}(\omega)=\sigma_0 \left[\frac{\sqrt{\tilde\omega^2-1}}{(\tilde\omega^2-1)+\xi^2\sin^2\gamma}\right]
\end{equation}
where $\xi=\beta^2/(1-\beta^2)$, $\gamma=\pi\left[1/2\beta^2-1\right]$, and $\tilde\omega =
\omega/2\Delta$ where $2\Delta$ is the semiconducting optical gap, and $\beta$ is the sine-Gordon
coupling constant \cite{controzzi01}.  When this functional form is taken in linear combination
with several Lorentzian oscillators, the optical conductivity is reproduced quite well with
$\sigma_0 = 1730$~$\Omega^{-1}$cm$^{-1}$, $2\Delta=614$~cm$^{-1}$, and $\beta=0.75$, as shown in
Fig.~3(b), clearly establishing the one-dimensional nature of the optical properties.
%
%
The estimate for $2\Delta$ along the \emph{b} axis considerably larger than $E_g$; however,
it should be noted that the optical determination of $2\Delta$ probes only direct transitions
between bands due to low momentum transfer.  If the material has a direct gap, then the optical
and transport gaps should be similar, $E_g \simeq 2\Delta$; however, in indirect-gap semiconductors,
phonon-assisted transitions typically result in $E_g < 2\Delta$.

\noindent The observation of one-dimensional behavior in this material is of particular importance
as it has been argued that lowered dimensionality may increase the value of the Seebeck
coefficient \cite{hicks93,kim09,rhyee09}.
%
%
\begin{figure}[t]
\includegraphics[width=3.4in]{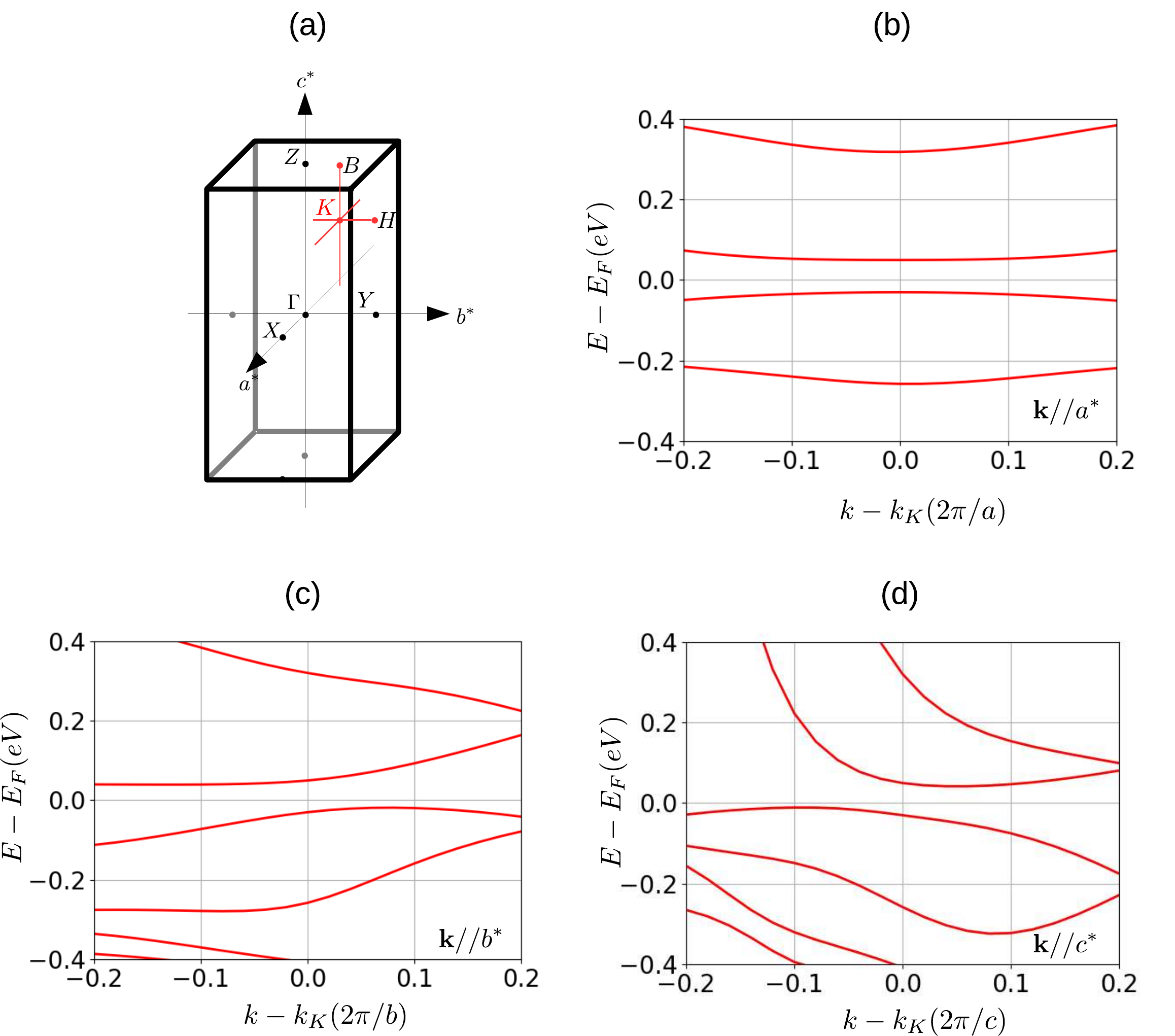}
\caption{Spectral function of FeSb$_{2}$.
(a) First Brillioun zone and high-symmetry points of FeSb$_2$ for the orthorhombic
phase.  Quasiparticle band structure near the K point ($0.26\mathbf{b}^\ast + 0.28\mathbf{c}^\ast$)
along the
(b) $\mathbf{a}^\ast$ direction; (c) $\mathbf{b}^\ast$ direction; (d), $\mathbf{c}^\ast$ direction.
The Brillouin zone cuts are indicated by the red lines in {\bf a}.}
\end{figure}
%
%
Electronic structure calculations can provide insight into the optical properties of a material.
However, density functional theory (DFT) predicts a metallic rather than a semiconducting ground
state, \cite{tomczak10} indicating that a more sophisticated approach is required.  Consequently,
first principle calculations have been performed using a linearized quasiparticle self-consistent
GW and dynamical mean field theory (LQSGW+DMFT) approach \cite{choi16,DMFT,kutepov17}
(details are provided in the Supplementary Information). Figure~4 shows the low-energy quasiparticle
band structure near the K point ($0.26\mathbf{b}^\ast + 0.28\mathbf{c}^\ast$) where the direct bandgap
is a minimum. Here $\mathbf{b}^\ast$ and $\mathbf{c}^\ast$ are the reciprocal lattice vectors
along the \emph{b} and \emph{c} axes.  Around the K point, the calculation shows direct bandgap
of $\simeq 80$~meV, which is in a good agreement with the semiconducting optical gap of
$2\Delta\simeq 76$~meV.  In addition, low-dimensional behavior is observed near the K point;
along the $\mathbf{a}^\ast$ direction, quasiparticle bands for the conduction and valence electrons
are almost flat, as illustrated by the quasiparticle band in Fig.~4(b).  In contrast, the quasiparticle
bands are dispersive along the $\mathbf{b}^\ast$ and $\mathbf{c}^\ast$ directions shown in
Figs.~4(c) and  (d).  The fact that DMFT is necessary to generate a low-dimensional
quasiparticle spectral function that is consistent with the semiconducting ground state
indicates that electronic correlations are an essential ingredient in understanding the
anisotropic optical and transport properties of FeSb$_2$.

%

%
%
\noindent We now turn our attention to the equally interesting behavior of the
infrared-active lattice modes.  FeSb$_2$ crystallizes in the orthorhombic $Pnnm$
space group, where \emph{c} is the short axis [Fig.~1(a)].  The irreducible vibrational
representation is then $\Gamma_{irr}=2A_g+2B_{1g}+B_{2g}+B_{3g}+2A_u+B_{1u}+3B_{2u}+3B_{3u}$,
of which only the $B_{1u}$, $B_{2u}$ and $B_{3u}$ modes are infrared-active along the
\emph{c}, \emph{b}, and \emph{a} axes, respectively \cite{perucchi06}.
The temperature dependence of the real part of the optical conductivity has been
projected onto the wave number versus temperature plane using the indicated color
scales in Figs.~5(a), (b), and (c) for light polarized along the \emph{a}, \emph{b},
and \emph{c} axes, respectively.  The vibrations have been fit using oscillators with
a Fano profile superimposed on an electronic background at 295 and 5~K (Supplementary
Figs.~S5, S6, and S7).  The frequencies of the lattice modes at the center of the
Brillouin zone and their atomic characters have also been calculated using first principles
techniques and are in good agreement with previous results \cite{miao12,lazarevic12} (details
are provided in the Supplementary Information); the comparison between theory and experiment
is shown in Table~1.

%
%
\begin{table}[t]
\caption{The experimentally-observed position ($\omega_j$), width ($\gamma_j$)
and strength ($\Omega_j$) of the infrared-active lattice modes in FeSb$_2$ along
the \emph{a} ($B_{3u}$), \emph{b} ($B_{2u}$), and \emph{c} ($B_{1u}$) axes at
295 and 5~K, compared with the frequencies and atomic intensities calculated
from first principles assuming a $Pnnm$ (orthorhombic) space group; for all of
the modes the asymmetry parameter $1/q_j^2\lesssim 0.01$ (symmetric profiles).
The phonon lifetimes $\tau_j \propto 1/\gamma_j$.  The uncertainties for the
fitted position, width, and strength are estimated to be 1\%, 5\%, and 10\%,
respectively.  All units are in cm$^{-1}$, unless otherwise indicated.}
\begin{ruledtabular}
\begin{tabular}{c ccc c ccc  c ccc}
%
          & \multicolumn{3}{c}{Theory} & & \multicolumn{7}{c}{Experiment} \\
          & & \multicolumn{2}{c}{Character} & & \multicolumn{3}{c}{295~K} & & \multicolumn{3}{c}{5~K} \\
 Mode     & $\omega_{calc}$ & Fe & Sb & & $\omega_j$ & $\gamma_j$ & $\Omega_j$
                                           & & $\omega_j$ & $\gamma_j$ & $\Omega_j$  \\
\hline
 $B_{1u}$ &     198  & 0.81 & 0.19 & & 191.0 &  4.2 & 1030 & & 195.1 & 1.0 & 1120 \\
\hline
 $B_{2u}$ &     112  & 0.09 & 0.91 & & 106.3 &  2.0 & 285 & & 110.2 & 0.9 &  267 \\
 $B_{2u}$ &     234  & 0.91 & 0.09 & & 231.0 &  3.4 & 608 & & 236.5 & 1.0 &  720 \\
 $B_{2u}$ &     284  & 0.81 & 0.19 & & 269.1 &  5.2 & 723 & & 276.9 & 1.3 &  900 \\
\hline
 $B_{3u}$ &     125  & 0.10 & 0.90 & & 120.8 &  2.1 & 203 & & 123.4 & 0.8 &  229 \\
          &     ---  & --- & ---  & & 191.3 & 12.9 & 355  & & ---   & --- &  --- \\
          &     ---  & --- & ---  & &  ---  &  --- & ---  & & 220.1 & 0.8 &  672 \\
 $B_{3u}$ &     252  & 0.98 & 0.02 & & 242.9 &  6.1 & 289 & & 251.2 & 3.0 &  130 \\
 $B_{3u}$ &     260  & 0.74 & 0.26 & & 253.7 &  5.6 & 743 & & 266.4 & 1.3 &  796 \\
%
%
\end{tabular}
\end{ruledtabular}
\end{table}

%
%
\begin{figure*}[t]
\includegraphics[width=6.8in]{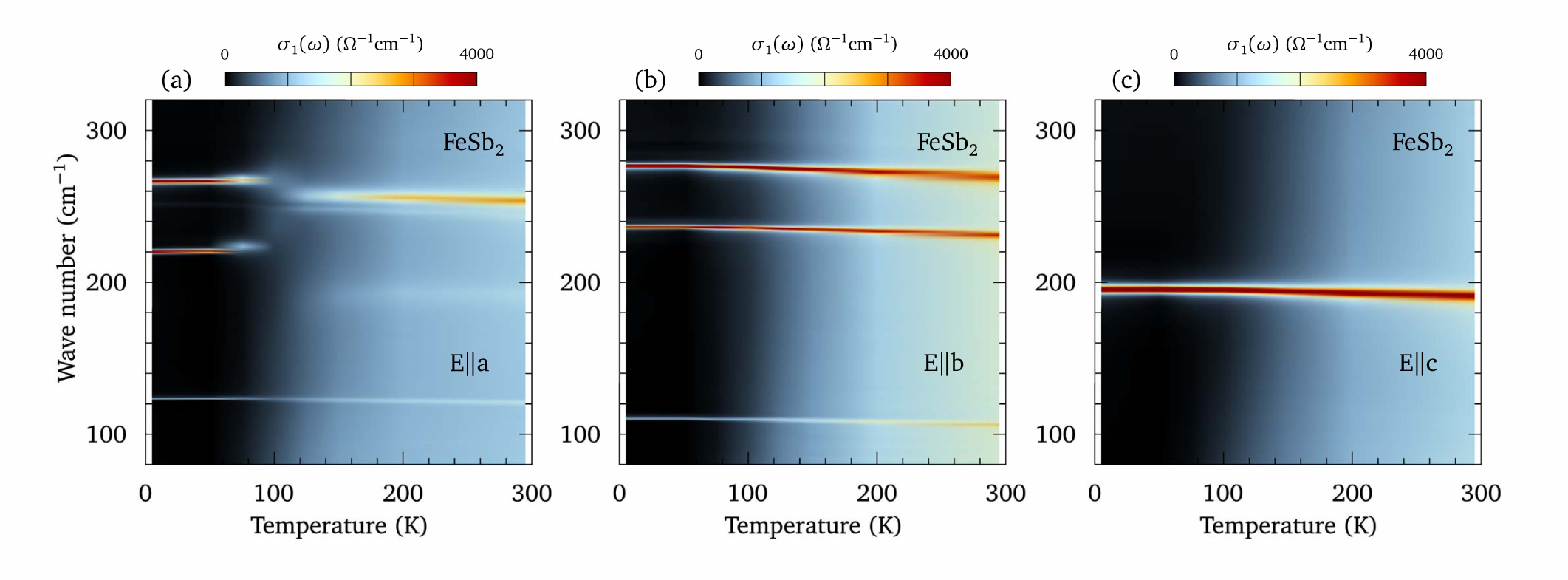}
\caption{Infrared-active phonons in FeSb$_{\mathbf 2}$.
(a) The temperature-dependence of the real part of the optical
conductivity for light polarized along the \emph{a} axis projected
onto the wave number versus temperature plane; the color scheme for
the conductivity is shown above the plot.  Only three $B_{3u}$ modes
are predicted for this symmetry; however, there are four modes at 121,
191, 243 and 254~cm$^{-1}$ at 295~K.  Below about 100~K the 191~cm$^{-1}$
mode disappears and is replaced by a new, very strong mode at
220~cm$^{-1}$; all the modes are quite narrow at low temperature
(Table 1).  The change in the character of the lattice modes
below 100~K hints at a weak structural distortion along this direction.
(b) The optical conductivity for light polarized along the
\emph{b} axis projected onto the wave number versus temperature plane.
There are three strong $B_{2u}$ modes at 106, 231, and 269~cm$^{-1}$
at 295~K that harden and narrow while increasingly slightly in strength
at low temperature.
(c) The optical conductivity for light polarized along the
\emph{c} axis projected onto the wave number versus temperature plane.
There is one strong $B_{1u}$ mode at 191~cm$^{-1}$ at 295~K that hardens
with decreasing temperature, increasing slightly in strength and narrowing
dramatically at low temperature.}
\end{figure*}

\noindent The behavior of the lattice modes are remarkable in several ways.  Along the
\emph{a}, \emph{b} and \emph{c} axes the vibrations have line widths that are up to an order
of magnitude smaller than the previously reported values \cite{perucchi06,herzog10};
at low temperature all the modes are extremely sharp and several have line widths of less
than 1~cm$^{-1}$, a result that has also been observed in some Raman-active
modes \cite{lazarevic10}.
The narrow line widths imply long phonon lifetimes ($\tau_j \propto 1/\gamma_j$) and
mean-free paths, consistent with the suggestion of quasi-ballistic
phonons \cite{takahashi16,battiato15}, which affect $S$ through the phonon-drag
effect where the phonon current drags the charge carriers, giving rise to an additional
thermoelectric voltage \cite{geballe54,herring54,weber91}.  In addition, while several
of the infrared-active vibrations were previously reported to have a slightly asymmetric
profile at high temperature \cite{perucchi06,herzog10}, in this work all the line shapes
appear to be symmetric ($1/q_j^2\simeq 0$), indicating that electron-phonon coupling is
either very weak or totally absent.
The single $B_{1u}$ mode along the \emph{c} axis, and the three $B_{2u}$ modes
along the \emph{b} axis, shown in Figs.~5(c) and (b), respectively, increase in
frequency (harden) with decreasing temperature, and are in excellent agreement with the
calculated values (Table~1).

The behavior of the lattice modes along the \emph{a} axis in Fig.~5(a) are
dramatically different.  At room temperature the three modes observed at
$\simeq 121$, 243 and 254~cm$^{-1}$ are in good agreement well with the calculated
values for the $B_{3u}$ modes at 125, 252, and 260~cm$^{-1}$, respectively;
however, a fourth reasonably strong mode at 191~cm$^{-1}$ is also observed that
is considerably broader than the other vibrations.  As the temperature is reduced the
mode at 191~cm$^{-1}$ actually decreases slightly in frequency, while the remaining
modes harden.
Below about 100~K, the mode at $\simeq 191$~cm$^{-1}$ vanishes and a new,
very strong mode appears at $\simeq 220$~cm$^{-1}$, while at the same time the
modes at 243 and 254~cm$^{-1}$ both shift to slightly higher frequencies;
the mode at 121~cm$^{-1}$ shows no signs of any anomalous behavior [Fig.~5(a)
and Fig.~S5].  The fate of the 191~cm$^{-1}$ mode is uncertain; however, it is unlikely
that it has evolved into the 220~cm$^{-1}$ mode due to the large difference in
oscillator strengths (Table~1).
It is also unlikely that this is a manifestation of the $B_{1u}$ mode, which
has a comparable frequency, because that feature does not display the unusual
temperature dependence of the mode observed along the \emph{a} axis, nor is
there any evidence of it along the \emph{b} axis.
The dramatic change in the nature of the lattice modes along the \emph{a} axis
at precisely the temperature where the resistivity begins to increase
dramatically suggests there is a weak structural distortion or phase transition.

%
%
{\em Summary.} To conclude, the temperature dependence of the optical and dc transport
properties of single crystals of FeSb$_2$, both with and without a MIT, have been
examined over a wide temperature and spectral range, along all three lattice
directions.  While the temperature dependence of the optical properties are
essentially identical in the two types of crystals, the dc transport properties
are dramatically different.  This dichotomy can be explained by the presence
of a sample-dependent impurity band that lies below the optical measurements.
The optical conductivity in both types of crystals reveals an anisotropic response at
room temperature, and singular behavior at low temperature along the $b$ axis,
demonstrating a one-dimensional semiconducting response with $2\Delta\simeq 76$~meV,
in agreement with \emph{ab inito} calculations.
The lattice modes along the $b$ and $c$ axes have symmetric profiles which
narrow and harden with decreasing temperature, and their positions are in
good agreement with first principles calculations.  However, along the \emph{a}
axis there is an extra mode above 100~K; below this temperature the resistivity
increases rapidly and the high-frequency vibrational modes undergo significant
changes that hint a weak structural distortion or transition.  Transport studies
along this direction may shed light on the nature of this peculiar behavior.
Although electron-phonon coupling is apparently either very weak or totally absent
in this material, the fact that DMFT is required to reproduce the semiconducting ground
state and anisotropic response indicates that electronic correlations
play an important role in the optical and transport properties.  While the extremely
narrow phonon line shapes support the phonon-drag explanation of the high thermoelectric
power, electronic correlations and the low-dimensional behavior along the \emph{b}
axis may also enhance the Seebeck coefficient \cite{hicks93,kim09,rhyee09}, making
it likely that both contribute to the extremely high thermopower observed
in FeSb$_2$.
%
%

%
%
{\em Methods.} The temperature dependence of the absolute reflectance was measured at a
near-normal angle of incidence using an {\em in situ} evaporation method \cite{homes93}
over a wide frequency range on Bruker IFS 113v and Vertex 80v spectrometers.  In this study
mirror-like as-grown faces of single crystals have been examined. After an initial
measurement, the \emph{c}-axis face was determined to have a minor surface
irregularity, so it was was polished and remeasured.  Polishing broadens
the lattice mode somewhat, but the electronic properties were not affected.
The temperature dependence of the reflectance was measured up to $\simeq 1.5$~eV,
while polarization studies were conducted up to at least 3~eV.
The complex optical properties were determined from a Kramers-Kronig analysis
of the reflectance \cite{dressel-book}.  The Kramers-Kronig transform requires
that the reflectance be determined for all frequencies, thus extrapolations
must be supplied in the $\omega \rightarrow 0, \infty$ limits.  In the metallic
state the low frequency extrapolation follows the Hagen-Rubens form,
$R(\omega) \propto 1-\sqrt{\omega}$, while in the semiconducting state the
reflectance was continued smoothly from the lowest measured frequency
point to $R(\omega\rightarrow 0)\simeq 0.64$ and 0.68 along the \emph{a} and
\emph{c} axes, respectively, and $\simeq 0.74$ along the \emph{b} axis.
The reflectance is assumed to be constant above the highest measured frequency
point up to $\simeq 8 \times 10^4$~cm$^{-1}$, above which a free electron gas
asymptotic reflectance extrapolation $R(\omega) \propto 1/\omega^4$ is
employed \cite{wooten}.

{\em Acknowledgments.}
The authors are grateful to T. Besara and T. Siegrist for confirming
the crystal orientation.  The authors would like to acknowledge useful discussions
with T. M. Rice, A. Tsvelik, T. Valla, R. Yang, and I. Zaliznyak.
Research supported by the U.S. Department of Energy, Office of
Basic Energy Sciences as part of the Computation Material Science Program through the
Center for Computational Material Spectroscopy and Design, and the Division of
Materials Sciences and Engineering under Contract No. DE-SC0012704.


%
%
%
%
%

%
%
\ \\

%
%
%

\begin{thebibliography}{35}%
\makeatletter
\providecommand \@ifxundefined [1]{%
 \@ifx{#1\undefined}
}%
\providecommand \@ifnum [1]{%
 \ifnum #1\expandafter \@firstoftwo
 \else \expandafter \@secondoftwo
 \fi
}%
\providecommand \@ifx [1]{%
 \ifx #1\expandafter \@firstoftwo
 \else \expandafter \@secondoftwo
 \fi
}%
\providecommand \natexlab [1]{#1}%
\providecommand \enquote  [1]{``#1''}%
\providecommand \bibnamefont  [1]{#1}%
\providecommand \bibfnamefont [1]{#1}%
\providecommand \citenamefont [1]{#1}%
\providecommand \href@noop [0]{\@secondoftwo}%
\providecommand \href [0]{\begingroup \@sanitize@url \@href}%
\providecommand \@href[1]{\@@startlink{#1}\@@href}%
\providecommand \@@href[1]{\endgroup#1\@@endlink}%
\providecommand \@sanitize@url [0]{\catcode `\\12\catcode `\$12\catcode
  `\&12\catcode `\#12\catcode `\^12\catcode `\_12\catcode `\%12\relax}%
\providecommand \@@startlink[1]{}%
\providecommand \@@endlink[0]{}%
\providecommand \url  [0]{\begingroup\@sanitize@url \@url }%
\providecommand \@url [1]{\endgroup\@href {#1}{\urlprefix }}%
\providecommand \urlprefix  [0]{URL }%
\providecommand \Eprint [0]{\href }%
\providecommand \doibase [0]{http://dx.doi.org/}%
\providecommand \selectlanguage [0]{\@gobble}%
\providecommand \bibinfo  [0]{\@secondoftwo}%
\providecommand \bibfield  [0]{\@secondoftwo}%
\providecommand \translation [1]{[#1]}%
\providecommand \BibitemOpen [0]{}%
\providecommand \bibitemStop [0]{}%
\providecommand \bibitemNoStop [0]{.\EOS\space}%
\providecommand \EOS [0]{\spacefactor3000\relax}%
\providecommand \BibitemShut  [1]{\csname bibitem#1\endcsname}%
\let\auto@bib@innerbib\@empty
\bibitem [{\citenamefont {Jie}\ \emph {et~al.}(2012)\citenamefont {Jie},
  \citenamefont {Hu}, \citenamefont {Bozin}, \citenamefont {Llobet},
  \citenamefont {Zaliznyak}, \citenamefont {Petrovic},\ and\ \citenamefont
  {Li}}]{jie12}%
  \BibitemOpen
  \bibfield  {author} {\bibinfo {author} {\bibfnamefont {Qing}\ \bibnamefont
  {Jie}}, \bibinfo {author} {\bibfnamefont {Rongwei}\ \bibnamefont {Hu}},
  \bibinfo {author} {\bibfnamefont {Emil}\ \bibnamefont {Bozin}}, \bibinfo
  {author} {\bibfnamefont {A.}~\bibnamefont {Llobet}}, \bibinfo {author}
  {\bibfnamefont {I.}~\bibnamefont {Zaliznyak}}, \bibinfo {author}
  {\bibfnamefont {C.}~\bibnamefont {Petrovic}}, \ and\ \bibinfo {author}
  {\bibfnamefont {Q.}~\bibnamefont {Li}},\ }\bibfield  {title} {\enquote
  {\bibinfo {title} {{Electronic thermoelectric power factor and
  metal-insulator transition in FeSb$_{2}$}},}\ }\href {\doibase
  10.1103/PhysRevB.86.115121} {\bibfield  {journal} {\bibinfo  {journal} {Phys.
  Rev. B}\ }\textbf {\bibinfo {volume} {86}},\ \bibinfo {pages} {115121}
  (\bibinfo {year} {2012})}\BibitemShut {NoStop}%
\bibitem [{\citenamefont {Bentien}\ \emph {et~al.}(2007)\citenamefont
  {Bentien}, \citenamefont {Johnsen}, \citenamefont {Madsen}, \citenamefont
  {Iversen},\ and\ \citenamefont {Steglich}}]{bentien07}%
  \BibitemOpen
  \bibfield  {author} {\bibinfo {author} {\bibfnamefont {A.}~\bibnamefont
  {Bentien}}, \bibinfo {author} {\bibfnamefont {S.}~\bibnamefont {Johnsen}},
  \bibinfo {author} {\bibfnamefont {G.~K.~H.}\ \bibnamefont {Madsen}}, \bibinfo
  {author} {\bibfnamefont {B.~B.}\ \bibnamefont {Iversen}}, \ and\ \bibinfo
  {author} {\bibfnamefont {F.}~\bibnamefont {Steglich}},\ }\bibfield  {title}
  {\enquote {\bibinfo {title} {{Colossal Seebeck coefficient in strongly
  correlated semiconductor FeSb$_2$}},}\ }\href {\doibase
  10.1209/0295-5075/80/17008} {\bibfield  {journal} {\bibinfo  {journal} {EPL}\
  }\textbf {\bibinfo {volume} {80}},\ \bibinfo {pages} {17008} (\bibinfo {year}
  {2007})}\BibitemShut {NoStop}%
\bibitem [{\citenamefont {Sun}\ \emph {et~al.}(2009)\citenamefont {Sun},
  \citenamefont {Oeschler}, \citenamefont {Johnsen}, \citenamefont {Iversen},\
  and\ \citenamefont {Steglich}}]{sun09}%
  \BibitemOpen
  \bibfield  {author} {\bibinfo {author} {\bibfnamefont {P}~\bibnamefont
  {Sun}}, \bibinfo {author} {\bibfnamefont {N}~\bibnamefont {Oeschler}},
  \bibinfo {author} {\bibfnamefont {S}~\bibnamefont {Johnsen}}, \bibinfo
  {author} {\bibfnamefont {B~B}\ \bibnamefont {Iversen}}, \ and\ \bibinfo
  {author} {\bibfnamefont {F}~\bibnamefont {Steglich}},\ }\bibfield  {title}
  {\enquote {\bibinfo {title} {{Thermoelectric properties of the narrow-gap
  semiconductors FeSb$_2$ and RuSb$_2$ : A comparative study}},}\ }\href
  {\doibase 1742-6596/150/i=1/a=012049} {\bibfield  {journal} {\bibinfo
  {journal} {Journal of Physics: Conference Series}\ }\textbf {\bibinfo
  {volume} {150}},\ \bibinfo {pages} {012049} (\bibinfo {year}
  {2009})}\BibitemShut {NoStop}%
\bibitem [{\citenamefont {Sun}\ \emph {et~al.}(2010)\citenamefont {Sun},
  \citenamefont {Oeschler}, \citenamefont {Johnsen}, \citenamefont {Iversen},\
  and\ \citenamefont {Steglich}}]{sun10}%
  \BibitemOpen
  \bibfield  {author} {\bibinfo {author} {\bibfnamefont {Peijie}\ \bibnamefont
  {Sun}}, \bibinfo {author} {\bibfnamefont {Niels}\ \bibnamefont {Oeschler}},
  \bibinfo {author} {\bibfnamefont {Simon}\ \bibnamefont {Johnsen}}, \bibinfo
  {author} {\bibfnamefont {Bo~B.}\ \bibnamefont {Iversen}}, \ and\ \bibinfo
  {author} {\bibfnamefont {Frank}\ \bibnamefont {Steglich}},\ }\bibfield
  {title} {\enquote {\bibinfo {title} {{Narrow band gap and enhanced
  thermoelectricity in FeSb$_2$}},}\ }\href {\doibase 10.1039/B918909B}
  {\bibfield  {journal} {\bibinfo  {journal} {Dalton Trans.}\ }\textbf
  {\bibinfo {volume} {39}},\ \bibinfo {pages} {1012--1019} (\bibinfo {year}
  {2010})}\BibitemShut {NoStop}%
\bibitem [{\citenamefont {Petrovic}\ \emph {et~al.}(2005)\citenamefont
  {Petrovic}, \citenamefont {Lee}, \citenamefont {Vogt}, \citenamefont
  {Lazarov}, \citenamefont {Bud'ko},\ and\ \citenamefont
  {Canfield}}]{petrovic05}%
  \BibitemOpen
  \bibfield  {author} {\bibinfo {author} {\bibfnamefont {C.}~\bibnamefont
  {Petrovic}}, \bibinfo {author} {\bibfnamefont {Y.}~\bibnamefont {Lee}},
  \bibinfo {author} {\bibfnamefont {T.}~\bibnamefont {Vogt}}, \bibinfo {author}
  {\bibfnamefont {N.~Dj.}\ \bibnamefont {Lazarov}}, \bibinfo {author}
  {\bibfnamefont {S.~L.}\ \bibnamefont {Bud'ko}}, \ and\ \bibinfo {author}
  {\bibfnamefont {P.~C.}\ \bibnamefont {Canfield}},\ }\bibfield  {title}
  {\enquote {\bibinfo {title} {Kondo insulator description of spin state
  transition in {FeSb}$_{2}$},}\ }\href {\doibase 10.1103/PhysRevB.72.045103}
  {\bibfield  {journal} {\bibinfo  {journal} {Phys. Rev. B}\ }\textbf {\bibinfo
  {volume} {72}},\ \bibinfo {pages} {045103} (\bibinfo {year}
  {2005})}\BibitemShut {NoStop}%
\bibitem [{\citenamefont {{Perucchi, A.}}\ \emph {et~al.}(2006)\citenamefont
  {{Perucchi, A.}}, \citenamefont {{Degiorgi, L.}}, \citenamefont {{Hu,
  Rongwei}}, \citenamefont {{Petrovic, C.}},\ and\ \citenamefont
  {{Mitrovi\'{c}, V. F.}}}]{perucchi06}%
  \BibitemOpen
  \bibfield  {author} {\bibinfo {author} {\bibnamefont {{Perucchi, A.}}},
  \bibinfo {author} {\bibnamefont {{Degiorgi, L.}}}, \bibinfo {author}
  {\bibnamefont {{Hu, Rongwei}}}, \bibinfo {author} {\bibnamefont {{Petrovic,
  C.}}}, \ and\ \bibinfo {author} {\bibnamefont {{Mitrovi\'{c}, V. F.}}},\
  }\bibfield  {title} {\enquote {\bibinfo {title} {{Optical investigation of
  the metal-insulator transition in FeSb$_2$}},}\ }\href {\doibase
  10.1140/epjb/e2006-00433-1} {\bibfield  {journal} {\bibinfo  {journal} {Eur.
  Phys. J. B}\ }\textbf {\bibinfo {volume} {54}},\ \bibinfo {pages} {175--183}
  (\bibinfo {year} {2006})}\BibitemShut {NoStop}%
\bibitem [{\citenamefont {Herzog}\ \emph {et~al.}(2010)\citenamefont {Herzog},
  \citenamefont {Marutzky}, \citenamefont {Sichelschmidt}, \citenamefont
  {Steglich}, \citenamefont {Kimura}, \citenamefont {Johnsen},\ and\
  \citenamefont {Iversen}}]{herzog10}%
  \BibitemOpen
  \bibfield  {author} {\bibinfo {author} {\bibfnamefont {A.}~\bibnamefont
  {Herzog}}, \bibinfo {author} {\bibfnamefont {M.}~\bibnamefont {Marutzky}},
  \bibinfo {author} {\bibfnamefont {J.}~\bibnamefont {Sichelschmidt}}, \bibinfo
  {author} {\bibfnamefont {F.}~\bibnamefont {Steglich}}, \bibinfo {author}
  {\bibfnamefont {S.}~\bibnamefont {Kimura}}, \bibinfo {author} {\bibfnamefont
  {S.}~\bibnamefont {Johnsen}}, \ and\ \bibinfo {author} {\bibfnamefont
  {B.~B.}\ \bibnamefont {Iversen}},\ }\bibfield  {title} {\enquote {\bibinfo
  {title} {{Strong electron correlations in {FeSb}$_{2}$: An optical
  investigation and comparison with {RuSb}$_{2}$}},}\ }\href {\doibase
  10.1103/PhysRevB.82.245205} {\bibfield  {journal} {\bibinfo  {journal} {Phys.
  Rev. B}\ }\textbf {\bibinfo {volume} {82}},\ \bibinfo {pages} {245205}
  (\bibinfo {year} {2010})}\BibitemShut {NoStop}%
\bibitem [{\citenamefont {{Figueira, M.S.}}\ \emph {et~al.}(2012)\citenamefont
  {{Figueira, M.S.}}, \citenamefont {{Silva-Valencia, J.}},\ and\ \citenamefont
  {{Franco, R.}}}]{figueira12}%
  \BibitemOpen
  \bibfield  {author} {\bibinfo {author} {\bibnamefont {{Figueira, M.S.}}},
  \bibinfo {author} {\bibnamefont {{Silva-Valencia, J.}}}, \ and\ \bibinfo
  {author} {\bibnamefont {{Franco, R.}}},\ }\bibfield  {title} {\enquote
  {\bibinfo {title} {{Thermoelectric properties of the Kondo insulator
  FeSb$_2$}},}\ }\href@noop {} {\bibfield  {journal} {\bibinfo  {journal} {Eur.
  Phys. J. B}\ }\textbf {\bibinfo {volume} {85}},\ \bibinfo {pages} {203}
  (\bibinfo {year} {2012})}\BibitemShut {NoStop}%
\bibitem [{\citenamefont {Fuccillo}\ \emph {et~al.}(2013)\citenamefont
  {Fuccillo}, \citenamefont {Gibson}, \citenamefont {Ali}, \citenamefont
  {Schoop},\ and\ \citenamefont {Cava}}]{fuccillo13}%
  \BibitemOpen
  \bibfield  {author} {\bibinfo {author} {\bibfnamefont {M.~K.}\ \bibnamefont
  {Fuccillo}}, \bibinfo {author} {\bibfnamefont {Q.~D.}\ \bibnamefont
  {Gibson}}, \bibinfo {author} {\bibfnamefont {Mazhar~N.}\ \bibnamefont {Ali}},
  \bibinfo {author} {\bibfnamefont {L.~M.}\ \bibnamefont {Schoop}}, \ and\
  \bibinfo {author} {\bibfnamefont {R.~J.}\ \bibnamefont {Cava}},\ }\bibfield
  {title} {\enquote {\bibinfo {title} {{Correlated evolution of colossal
  thermoelectric effect and Kondo insulating behavior}},}\ }\href {\doibase
  10.1063/1.4833055} {\bibfield  {journal} {\bibinfo  {journal} {APL Mater.}\
  }\textbf {\bibinfo {volume} {1}},\ \bibinfo {pages} {062102} (\bibinfo {year}
  {2013})}\BibitemShut {NoStop}%
\bibitem [{\citenamefont {Sun}\ \emph {et~al.}(2013)\citenamefont {Sun},
  \citenamefont {Xu}, \citenamefont {Tomczak}, \citenamefont {Kotliar},
  \citenamefont {S\o{}ndergaard}, \citenamefont {Iversen},\ and\ \citenamefont
  {Steglich}}]{sun13}%
  \BibitemOpen
  \bibfield  {author} {\bibinfo {author} {\bibfnamefont {Peijie}\ \bibnamefont
  {Sun}}, \bibinfo {author} {\bibfnamefont {Wenhu}\ \bibnamefont {Xu}},
  \bibinfo {author} {\bibfnamefont {Jan~M.}\ \bibnamefont {Tomczak}}, \bibinfo
  {author} {\bibfnamefont {Gabriel}\ \bibnamefont {Kotliar}}, \bibinfo {author}
  {\bibfnamefont {Martin}\ \bibnamefont {S\o{}ndergaard}}, \bibinfo {author}
  {\bibfnamefont {Bo~B.}\ \bibnamefont {Iversen}}, \ and\ \bibinfo {author}
  {\bibfnamefont {Frank}\ \bibnamefont {Steglich}},\ }\bibfield  {title}
  {\enquote {\bibinfo {title} {{Highly dispersive electron relaxation and
  colossal thermoelectricity in the correlated semiconductor FeSb$_{2}$}},}\
  }\href {\doibase 10.1103/PhysRevB.88.245203} {\bibfield  {journal} {\bibinfo
  {journal} {Phys. Rev. B}\ }\textbf {\bibinfo {volume} {88}},\ \bibinfo
  {pages} {245203} (\bibinfo {year} {2013})}\BibitemShut {NoStop}%
\bibitem [{\citenamefont {Takahashi}\ \emph {et~al.}(2016)\citenamefont
  {Takahashi}, \citenamefont {Okazaki}, \citenamefont {Ishiwata}, \citenamefont
  {Taniguchi}, \citenamefont {Okutani}, \citenamefont {Hagiwara},\ and\
  \citenamefont {Terasaki}}]{takahashi16}%
  \BibitemOpen
  \bibfield  {author} {\bibinfo {author} {\bibfnamefont {H.}~\bibnamefont
  {Takahashi}}, \bibinfo {author} {\bibfnamefont {R.}~\bibnamefont {Okazaki}},
  \bibinfo {author} {\bibfnamefont {S.}~\bibnamefont {Ishiwata}}, \bibinfo
  {author} {\bibfnamefont {H.}~\bibnamefont {Taniguchi}}, \bibinfo {author}
  {\bibfnamefont {A.}~\bibnamefont {Okutani}}, \bibinfo {author} {\bibfnamefont
  {M.}~\bibnamefont {Hagiwara}}, \ and\ \bibinfo {author} {\bibfnamefont
  {I.}~\bibnamefont {Terasaki}},\ }\bibfield  {title} {\enquote {\bibinfo
  {title} {{Colossal Seebeck effect enhanced by quasi-ballistic phonons
  dragging massive electrons in FeSb$_2$}},}\ }\href {\doibase
  10.1038/ncomms12732} {\bibfield  {journal} {\bibinfo  {journal} {Nature
  Commun.}\ }\textbf {\bibinfo {volume} {7}},\ \bibinfo {pages} {12732}
  (\bibinfo {year} {2016})}\BibitemShut {NoStop}%
\bibitem [{\citenamefont {Tomczak}\ \emph {et~al.}(2010)\citenamefont
  {Tomczak}, \citenamefont {Haule}, \citenamefont {Miyake}, \citenamefont
  {Georges},\ and\ \citenamefont {Kotliar}}]{tomczak10}%
  \BibitemOpen
  \bibfield  {author} {\bibinfo {author} {\bibfnamefont {Jan~M.}\ \bibnamefont
  {Tomczak}}, \bibinfo {author} {\bibfnamefont {K.}~\bibnamefont {Haule}},
  \bibinfo {author} {\bibfnamefont {T.}~\bibnamefont {Miyake}}, \bibinfo
  {author} {\bibfnamefont {A.}~\bibnamefont {Georges}}, \ and\ \bibinfo
  {author} {\bibfnamefont {G.}~\bibnamefont {Kotliar}},\ }\bibfield  {title}
  {\enquote {\bibinfo {title} {{Thermopower of correlated semiconductors:
  Application to ${\text{FeAs}}_{2}$ and ${\text{FeSb}}_{2}$}},}\ }\href
  {\doibase 10.1103/PhysRevB.82.085104} {\bibfield  {journal} {\bibinfo
  {journal} {Phys. Rev. B}\ }\textbf {\bibinfo {volume} {82}},\ \bibinfo
  {pages} {085104} (\bibinfo {year} {2010})}\BibitemShut {NoStop}%
\bibitem [{\citenamefont {Pokharel}\ \emph {et~al.}(2013)\citenamefont
  {Pokharel}, \citenamefont {Zhao}, \citenamefont {Lukas}, \citenamefont {Ren},
  \citenamefont {Opeil},\ and\ \citenamefont {Mihaila}}]{pokharel13}%
  \BibitemOpen
  \bibfield  {author} {\bibinfo {author} {\bibfnamefont {Mani}\ \bibnamefont
  {Pokharel}}, \bibinfo {author} {\bibfnamefont {Huaizhou}\ \bibnamefont
  {Zhao}}, \bibinfo {author} {\bibfnamefont {Kevin}\ \bibnamefont {Lukas}},
  \bibinfo {author} {\bibfnamefont {Zhifeng}\ \bibnamefont {Ren}}, \bibinfo
  {author} {\bibfnamefont {Cyril}\ \bibnamefont {Opeil}}, \ and\ \bibinfo
  {author} {\bibfnamefont {Bogdan}\ \bibnamefont {Mihaila}},\ }\bibfield
  {title} {\enquote {\bibinfo {title} {{Phonon drag effect in nanocomposite
  FeSb$_2$}},}\ }\href {\doibase 10.1557/mrc.2013.7} {\bibfield  {journal}
  {\bibinfo  {journal} {MRS Commun.}\ }\textbf {\bibinfo {volume} {3}},\
  \bibinfo {pages} {31--36} (\bibinfo {year} {2013})}\BibitemShut {NoStop}%
\bibitem [{\citenamefont {Liao}\ \emph {et~al.}(2014)\citenamefont {Liao},
  \citenamefont {Lee}, \citenamefont {Esfarjani},\ and\ \citenamefont
  {Chen}}]{liao14}%
  \BibitemOpen
  \bibfield  {author} {\bibinfo {author} {\bibfnamefont {Bolin}\ \bibnamefont
  {Liao}}, \bibinfo {author} {\bibfnamefont {Sangyeop}\ \bibnamefont {Lee}},
  \bibinfo {author} {\bibfnamefont {Keivan}\ \bibnamefont {Esfarjani}}, \ and\
  \bibinfo {author} {\bibfnamefont {Gang}\ \bibnamefont {Chen}},\ }\bibfield
  {title} {\enquote {\bibinfo {title} {{First-principles study of thermal
  transport in {FeSb}$_{2}$}},}\ }\href {\doibase 10.1103/PhysRevB.89.035108}
  {\bibfield  {journal} {\bibinfo  {journal} {Phys. Rev. B}\ }\textbf {\bibinfo
  {volume} {89}},\ \bibinfo {pages} {035108} (\bibinfo {year}
  {2014})}\BibitemShut {NoStop}%
\bibitem [{\citenamefont {Battiato}\ \emph {et~al.}(2015)\citenamefont
  {Battiato}, \citenamefont {Tomczak}, \citenamefont {Zhong},\ and\
  \citenamefont {Held}}]{battiato15}%
  \BibitemOpen
  \bibfield  {author} {\bibinfo {author} {\bibfnamefont {M.}~\bibnamefont
  {Battiato}}, \bibinfo {author} {\bibfnamefont {J.~M.}\ \bibnamefont
  {Tomczak}}, \bibinfo {author} {\bibfnamefont {Z.}~\bibnamefont {Zhong}}, \
  and\ \bibinfo {author} {\bibfnamefont {K.}~\bibnamefont {Held}},\ }\bibfield
  {title} {\enquote {\bibinfo {title} {{Unified Picture for the Colossal
  Thermopower Compound ${\mathrm{FeSb}}_{2}$}},}\ }\href {\doibase
  10.1103/PhysRevLett.114.236603} {\bibfield  {journal} {\bibinfo  {journal}
  {Phys. Rev. Lett.}\ }\textbf {\bibinfo {volume} {114}},\ \bibinfo {pages}
  {236603} (\bibinfo {year} {2015})}\BibitemShut {NoStop}%
\bibitem [{\citenamefont {Holseth}\ and\ \citenamefont
  {Kjekshus}(1968)}]{holseth68}%
  \BibitemOpen
  \bibfield  {author} {\bibinfo {author} {\bibfnamefont {Hans}\ \bibnamefont
  {Holseth}}\ and\ \bibinfo {author} {\bibfnamefont {Arne}\ \bibnamefont
  {Kjekshus}},\ }\bibfield  {title} {\enquote {\bibinfo {title} {{Compounds
  with the Marcasite Type Crystal Structure. IV. The Crystal Structure of
  FeSb$_2$}},}\ }\href {\doibase 10.3891/acta.chem.scand.22-3284} {\bibfield
  {journal} {\bibinfo  {journal} {Acta Chem. Scand.}\ }\textbf {\bibinfo
  {volume} {23}},\ \bibinfo {pages} {3043--3050} (\bibinfo {year}
  {1968})}\BibitemShut {NoStop}%
\bibitem [{\citenamefont {Petrovic}\ \emph {et~al.}(2003)\citenamefont
  {Petrovic}, \citenamefont {Kim}, \citenamefont {Bud'ko}, \citenamefont
  {Goldman}, \citenamefont {Canfield}, \citenamefont {Choe},\ and\
  \citenamefont {Miller}}]{petrovic03}%
  \BibitemOpen
  \bibfield  {author} {\bibinfo {author} {\bibfnamefont {C.}~\bibnamefont
  {Petrovic}}, \bibinfo {author} {\bibfnamefont {J.~W.}\ \bibnamefont {Kim}},
  \bibinfo {author} {\bibfnamefont {S.~L.}\ \bibnamefont {Bud'ko}}, \bibinfo
  {author} {\bibfnamefont {A.~I.}\ \bibnamefont {Goldman}}, \bibinfo {author}
  {\bibfnamefont {P.~C.}\ \bibnamefont {Canfield}}, \bibinfo {author}
  {\bibfnamefont {W.}~\bibnamefont {Choe}}, \ and\ \bibinfo {author}
  {\bibfnamefont {G.~J.}\ \bibnamefont {Miller}},\ }\bibfield  {title}
  {\enquote {\bibinfo {title} {{Anisotropy and large magnetoresistance in the
  narrow-gap semiconductor {FeSb}$_{2}$}},}\ }\href {\doibase
  10.1103/PhysRevB.67.155205} {\bibfield  {journal} {\bibinfo  {journal} {Phys.
  Rev. B}\ }\textbf {\bibinfo {volume} {67}},\ \bibinfo {pages} {155205}
  (\bibinfo {year} {2003})}\BibitemShut {NoStop}%
\bibitem [{\citenamefont {Bentien}\ \emph {et~al.}(2006)\citenamefont
  {Bentien}, \citenamefont {Madsen}, \citenamefont {Johnsen},\ and\
  \citenamefont {Iversen}}]{bentien06}%
  \BibitemOpen
  \bibfield  {author} {\bibinfo {author} {\bibfnamefont {A.}~\bibnamefont
  {Bentien}}, \bibinfo {author} {\bibfnamefont {G.~K.~H.}\ \bibnamefont
  {Madsen}}, \bibinfo {author} {\bibfnamefont {S.}~\bibnamefont {Johnsen}}, \
  and\ \bibinfo {author} {\bibfnamefont {B.~B.}\ \bibnamefont {Iversen}},\
  }\bibfield  {title} {\enquote {\bibinfo {title} {{Experimental and
  theoretical investigations of strongly correlated {FeSb}$_{2-x}${Sn}$_x$}},}\
  }\href {\doibase 10.1103/PhysRevB.74.205105} {\bibfield  {journal} {\bibinfo
  {journal} {Phys. Rev. B}\ }\textbf {\bibinfo {volume} {74}},\ \bibinfo
  {pages} {205105} (\bibinfo {year} {2006})}\BibitemShut {NoStop}%
\bibitem [{\citenamefont {Homes}\ \emph {et~al.}(1993)\citenamefont {Homes},
  \citenamefont {Reedyk}, \citenamefont {Crandles},\ and\ \citenamefont
  {Timusk}}]{homes93}%
  \BibitemOpen
  \bibfield  {author} {\bibinfo {author} {\bibfnamefont {Christopher~C.}\
  \bibnamefont {Homes}}, \bibinfo {author} {\bibfnamefont {M.}~\bibnamefont
  {Reedyk}}, \bibinfo {author} {\bibfnamefont {D.~A.}\ \bibnamefont
  {Crandles}}, \ and\ \bibinfo {author} {\bibfnamefont {T.}~\bibnamefont
  {Timusk}},\ }\bibfield  {title} {\enquote {\bibinfo {title} {{Technique for
  measuring the reflectance of irregular, submillimeter-sized samples}},}\
  }\href {\doibase 10.1364/AO.32.002976} {\bibfield  {journal} {\bibinfo
  {journal} {Appl. Opt.}\ }\textbf {\bibinfo {volume} {32}},\ \bibinfo {pages}
  {2976--2983} (\bibinfo {year} {1993})}\BibitemShut {NoStop}%
\bibitem [{\citenamefont {Dressel}\ and\ \citenamefont
  {Gr{\"u}ner}(2001)}]{dressel-book}%
  \BibitemOpen
  \bibfield  {author} {\bibinfo {author} {\bibfnamefont {M.}~\bibnamefont
  {Dressel}}\ and\ \bibinfo {author} {\bibfnamefont {G.}~\bibnamefont
  {Gr{\"u}ner}},\ }\href@noop {} {\emph {\bibinfo {title} {Electrodynamics of
  Solids}}}\ (\bibinfo  {publisher} {Cambridge University Press},\ \bibinfo
  {address} {Cambridge},\ \bibinfo {year} {2001})\BibitemShut {NoStop}%
\bibitem [{\citenamefont {Homes}\ \emph {et~al.}(2016)\citenamefont {Homes},
  \citenamefont {Dai}, \citenamefont {Schneeloch}, \citenamefont {Zhong},\ and\
  \citenamefont {Gu}}]{homes16}%
  \BibitemOpen
  \bibfield  {author} {\bibinfo {author} {\bibfnamefont {C.~C.}\ \bibnamefont
  {Homes}}, \bibinfo {author} {\bibfnamefont {Y.~M.}\ \bibnamefont {Dai}},
  \bibinfo {author} {\bibfnamefont {J.}~\bibnamefont {Schneeloch}}, \bibinfo
  {author} {\bibfnamefont {R.~D.}\ \bibnamefont {Zhong}}, \ and\ \bibinfo
  {author} {\bibfnamefont {G.~D.}\ \bibnamefont {Gu}},\ }\bibfield  {title}
  {\enquote {\bibinfo {title} {Phonon anomalies in some iron telluride
  materials},}\ }\href {\doibase 10.1103/PhysRevB.93.125135} {\bibfield
  {journal} {\bibinfo  {journal} {Phys. Rev. B}\ }\textbf {\bibinfo {volume}
  {93}},\ \bibinfo {pages} {125135} (\bibinfo {year} {2016})}\BibitemShut
  {NoStop}%
\bibitem [{\citenamefont {Controzzi}\ \emph {et~al.}(2001)\citenamefont
  {Controzzi}, \citenamefont {Essler},\ and\ \citenamefont
  {Tsvelik}}]{controzzi01}%
  \BibitemOpen
  \bibfield  {author} {\bibinfo {author} {\bibfnamefont {D.}~\bibnamefont
  {Controzzi}}, \bibinfo {author} {\bibfnamefont {F.~H.~L.}\ \bibnamefont
  {Essler}}, \ and\ \bibinfo {author} {\bibfnamefont {A.~M.}\ \bibnamefont
  {Tsvelik}},\ }\bibfield  {title} {\enquote {\bibinfo {title} {{Optical
  Conductivity of One-Dimensional Mott Insulators}},}\ }\href {\doibase
  10.1103/PhysRevLett.86.680} {\bibfield  {journal} {\bibinfo  {journal} {Phys.
  Rev. Lett.}\ }\textbf {\bibinfo {volume} {86}},\ \bibinfo {pages} {680--683}
  (\bibinfo {year} {2001})}\BibitemShut {NoStop}%
\bibitem [{\citenamefont {Hicks}\ and\ \citenamefont
  {Dresselhaus}(1993)}]{hicks93}%
  \BibitemOpen
  \bibfield  {author} {\bibinfo {author} {\bibfnamefont {L.~D.}\ \bibnamefont
  {Hicks}}\ and\ \bibinfo {author} {\bibfnamefont {M.~S.}\ \bibnamefont
  {Dresselhaus}},\ }\bibfield  {title} {\enquote {\bibinfo {title}
  {Thermoelectric figure of merit of a one-dimensional conductor},}\ }\href
  {\doibase 10.1103/PhysRevB.47.16631} {\bibfield  {journal} {\bibinfo
  {journal} {Phys. Rev. B}\ }\textbf {\bibinfo {volume} {47}},\ \bibinfo
  {pages} {16631(R)} (\bibinfo {year} {1993})}\BibitemShut {NoStop}%
\bibitem [{\citenamefont {Kim}\ \emph {et~al.}(2009)\citenamefont {Kim},
  \citenamefont {Datta},\ and\ \citenamefont {Lundstrom}}]{kim09}%
  \BibitemOpen
  \bibfield  {author} {\bibinfo {author} {\bibfnamefont {Raseong}\ \bibnamefont
  {Kim}}, \bibinfo {author} {\bibfnamefont {Supriyo}\ \bibnamefont {Datta}}, \
  and\ \bibinfo {author} {\bibfnamefont {Mark~S.}\ \bibnamefont {Lundstrom}},\
  }\bibfield  {title} {\enquote {\bibinfo {title} {Influence of dimensionality
  on thermoelectric device performance},}\ }\href {\doibase 10.1063/1.3074347}
  {\bibfield  {journal} {\bibinfo  {journal} {J. App. Phys.}\ }\textbf
  {\bibinfo {volume} {105}},\ \bibinfo {pages} {034506} (\bibinfo {year}
  {2009})}\BibitemShut {NoStop}%
\bibitem [{\citenamefont {Rhyee}\ \emph {et~al.}(2009)\citenamefont {Rhyee},
  \citenamefont {Lee}, \citenamefont {Lee}, \citenamefont {Cho}, \citenamefont
  {Kim}, \citenamefont {Lee}, \citenamefont {Kwon}, \citenamefont {Shim},\ and\
  \citenamefont {Kotliar}}]{rhyee09}%
  \BibitemOpen
  \bibfield  {author} {\bibinfo {author} {\bibfnamefont {Jong-Soo}\
  \bibnamefont {Rhyee}}, \bibinfo {author} {\bibfnamefont {Kyu~Hyoung}\
  \bibnamefont {Lee}}, \bibinfo {author} {\bibfnamefont {Sang~Mock}\
  \bibnamefont {Lee}}, \bibinfo {author} {\bibfnamefont {Eunseog}\ \bibnamefont
  {Cho}}, \bibinfo {author} {\bibfnamefont {Sang~Il}\ \bibnamefont {Kim}},
  \bibinfo {author} {\bibfnamefont {Eunsung}\ \bibnamefont {Lee}}, \bibinfo
  {author} {\bibfnamefont {Yong~Seung}\ \bibnamefont {Kwon}}, \bibinfo {author}
  {\bibfnamefont {Ji~Hoon}\ \bibnamefont {Shim}}, \ and\ \bibinfo {author}
  {\bibfnamefont {Gabriel}\ \bibnamefont {Kotliar}},\ }\bibfield  {title}
  {\enquote {\bibinfo {title} {{Peierls distortion as a route to high
  thermoelectric performance in In$_4$Se$_{3-\delta}$ crystals}},}\ }\href
  {\doibase 10.1038/nature08088} {\bibfield  {journal} {\bibinfo  {journal}
  {Nature}\ }\textbf {\bibinfo {volume} {459}},\ \bibinfo {pages} {965--968}
  (\bibinfo {year} {2009})}\BibitemShut {NoStop}%
\bibitem [{\citenamefont {Choi}\ \emph {et~al.}(2016)\citenamefont {Choi},
  \citenamefont {Kutepov}, \citenamefont {Haule}, \citenamefont {{van
  Schilfgaarde}},\ and\ \citenamefont {Kotliar}}]{choi16}%
  \BibitemOpen
  \bibfield  {author} {\bibinfo {author} {\bibfnamefont {Sangkook}\
  \bibnamefont {Choi}}, \bibinfo {author} {\bibfnamefont {Andrey}\ \bibnamefont
  {Kutepov}}, \bibinfo {author} {\bibfnamefont {Kristjan}\ \bibnamefont
  {Haule}}, \bibinfo {author} {\bibfnamefont {Mark}\ \bibnamefont {{van
  Schilfgaarde}}}, \ and\ \bibinfo {author} {\bibfnamefont {Gabriel}\
  \bibnamefont {Kotliar}},\ }\bibfield  {title} {\enquote {\bibinfo {title}
  {{First-principles treatment of Mott insulators: linearized QSGW+DMFT
  approach}},}\ }\href {\doibase 10.1038/npjquantmats.2016.1} {\bibfield
  {journal} {\bibinfo  {journal} {Quantum Materials}\ }\textbf {\bibinfo
  {volume} {1}},\ \bibinfo {pages} {16001} (\bibinfo {year}
  {2016})}\BibitemShut {NoStop}%
\bibitem [{DMF()}]{DMFT}%
  \BibitemOpen
  \href@noop {} {}\bibinfo {note} {{For the GW part of the LQSGW+DMFT scheme,
  the code FlapwMBPT was used (http://scgw.physics.rutgers.edu).}}\BibitemShut
  {Stop}%
\bibitem [{\citenamefont {Kutepov}\ \emph {et~al.}(2017)\citenamefont
  {Kutepov}, \citenamefont {Oudovenko},\ and\ \citenamefont
  {Kotliar}}]{kutepov17}%
  \BibitemOpen
  \bibfield  {author} {\bibinfo {author} {\bibfnamefont {A.~L.}\ \bibnamefont
  {Kutepov}}, \bibinfo {author} {\bibfnamefont {V.~S.}\ \bibnamefont
  {Oudovenko}}, \ and\ \bibinfo {author} {\bibfnamefont {G.}~\bibnamefont
  {Kotliar}},\ }\bibfield  {title} {\enquote {\bibinfo {title} {{Linearized
  self-consistent quasiparticle GW method: Application to semiconductors and
  simple metals}},}\ }\href {\doibase 10.1016/j.cpc.2017.06.012} {\bibfield
  {journal} {\bibinfo  {journal} {Comp. Phys. Commun.}\ }\textbf {\bibinfo
  {volume} {219}},\ \bibinfo {pages} {407--414} (\bibinfo {year}
  {2017})}\BibitemShut {NoStop}%
\bibitem [{\citenamefont {Miao}\ \emph {et~al.}(2012)\citenamefont {Miao},
  \citenamefont {Huang}, \citenamefont {Fan}, \citenamefont {Bai},
  \citenamefont {Li}, \citenamefont {Wang}, \citenamefont {Chen}, \citenamefont
  {Song},\ and\ \citenamefont {Xu}}]{miao12}%
  \BibitemOpen
  \bibfield  {author} {\bibinfo {author} {\bibfnamefont {Rende}\ \bibnamefont
  {Miao}}, \bibinfo {author} {\bibfnamefont {Guiqin}\ \bibnamefont {Huang}},
  \bibinfo {author} {\bibfnamefont {Chunhui}\ \bibnamefont {Fan}}, \bibinfo
  {author} {\bibfnamefont {Zhong}\ \bibnamefont {Bai}}, \bibinfo {author}
  {\bibfnamefont {Yanbiao}\ \bibnamefont {Li}}, \bibinfo {author}
  {\bibfnamefont {Liang}\ \bibnamefont {Wang}}, \bibinfo {author}
  {\bibfnamefont {{Li an}}\ \bibnamefont {Chen}}, \bibinfo {author}
  {\bibfnamefont {Wenguang}\ \bibnamefont {Song}}, \ and\ \bibinfo {author}
  {\bibfnamefont {Qiangui}\ \bibnamefont {Xu}},\ }\bibfield  {title} {\enquote
  {\bibinfo {title} {{First-principles study on the lattice dynamics of
  FeSb$_2$}},}\ }\href {\doibase 10.1016/j.ssc.2011.10.022} {\bibfield
  {journal} {\bibinfo  {journal} {Solid State Commun.}\ }\textbf {\bibinfo
  {volume} {152}},\ \bibinfo {pages} {231--234} (\bibinfo {year}
  {2012})}\BibitemShut {NoStop}%
\bibitem [{\citenamefont {Lazarevi\'{c}}\ \emph {et~al.}(2012)\citenamefont
  {Lazarevi\'{c}}, \citenamefont {Radonji\'{c}}, \citenamefont
  {Tanaskovi\'{c}}, \citenamefont {Hu}, \citenamefont {Petrovic},\ and\
  \citenamefont {Popovi\'{c}}}]{lazarevic12}%
  \BibitemOpen
  \bibfield  {author} {\bibinfo {author} {\bibfnamefont {N.}~\bibnamefont
  {Lazarevi\'{c}}}, \bibinfo {author} {\bibfnamefont {M.~M.}\ \bibnamefont
  {Radonji\'{c}}}, \bibinfo {author} {\bibfnamefont {D.}~\bibnamefont
  {Tanaskovi\'{c}}}, \bibinfo {author} {\bibfnamefont {Rongwei}\ \bibnamefont
  {Hu}}, \bibinfo {author} {\bibfnamefont {C.}~\bibnamefont {Petrovic}}, \ and\
  \bibinfo {author} {\bibfnamefont {Z.~V.}\ \bibnamefont {Popovi\'{c}}},\
  }\bibfield  {title} {\enquote {\bibinfo {title} {{Lattice dynamics of
  FeSb$_2$}},}\ }\href {\doibase 10.1088/0953-8984/24/25/255402} {\bibfield
  {journal} {\bibinfo  {journal} {J. Phys.: Condens. Matter}\ }\textbf
  {\bibinfo {volume} {24}},\ \bibinfo {pages} {255402} (\bibinfo {year}
  {2012})}\BibitemShut {NoStop}%
\bibitem [{\citenamefont {Lazarevi\ifmmode~\acute{c}\else \'{c}\fi{}}\ \emph
  {et~al.}(2010)\citenamefont {Lazarevi\ifmmode~\acute{c}\else \'{c}\fi{}},
  \citenamefont {Popovi\ifmmode~\acute{c}\else \'{c}\fi{}}, \citenamefont
  {Hu},\ and\ \citenamefont {Petrovic}}]{lazarevic10}%
  \BibitemOpen
  \bibfield  {author} {\bibinfo {author} {\bibfnamefont {N.}~\bibnamefont
  {Lazarevi\ifmmode~\acute{c}\else \'{c}\fi{}}}, \bibinfo {author}
  {\bibfnamefont {Z.~V.}\ \bibnamefont {Popovi\ifmmode~\acute{c}\else
  \'{c}\fi{}}}, \bibinfo {author} {\bibfnamefont {Rongwei}\ \bibnamefont {Hu}},
  \ and\ \bibinfo {author} {\bibfnamefont {C.}~\bibnamefont {Petrovic}},\
  }\bibfield  {title} {\enquote {\bibinfo {title} {{Evidence for
  electron-phonon interaction in {Fe}$_{1-x}{M}_{x}${Sb}$_{2}$ ($M=\,$ Co and
  Cr; $0\ensuremath{\le}x\ensuremath{\le}0.5$) single crystals}},}\ }\href
  {\doibase 10.1103/PhysRevB.81.144302} {\bibfield  {journal} {\bibinfo
  {journal} {Phys. Rev. B}\ }\textbf {\bibinfo {volume} {81}},\ \bibinfo
  {pages} {144302} (\bibinfo {year} {2010})}\BibitemShut {NoStop}%
\bibitem [{\citenamefont {Geballe}\ and\ \citenamefont
  {Hull}(1954)}]{geballe54}%
  \BibitemOpen
  \bibfield  {author} {\bibinfo {author} {\bibfnamefont {T.~H.}\ \bibnamefont
  {Geballe}}\ and\ \bibinfo {author} {\bibfnamefont {G.~W.}\ \bibnamefont
  {Hull}},\ }\bibfield  {title} {\enquote {\bibinfo {title} {{Seebeck Effect in
  Germanium}},}\ }\href {\doibase 10.1103/PhysRev.94.1134} {\bibfield
  {journal} {\bibinfo  {journal} {Phys. Rev.}\ }\textbf {\bibinfo {volume}
  {94}},\ \bibinfo {pages} {1134--1140} (\bibinfo {year} {1954})}\BibitemShut
  {NoStop}%
\bibitem [{\citenamefont {Herring}(1954)}]{herring54}%
  \BibitemOpen
  \bibfield  {author} {\bibinfo {author} {\bibfnamefont {Conyers}\ \bibnamefont
  {Herring}},\ }\bibfield  {title} {\enquote {\bibinfo {title} {{Theory of the
  Thermoelectric Power of Semiconductors}},}\ }\href {\doibase
  10.1103/PhysRev.96.1163} {\bibfield  {journal} {\bibinfo  {journal} {Phys.
  Rev.}\ }\textbf {\bibinfo {volume} {96}},\ \bibinfo {pages} {1163--1187}
  (\bibinfo {year} {1954})}\BibitemShut {NoStop}%
\bibitem [{\citenamefont {Weber}\ and\ \citenamefont {Gmelin}(1991)}]{weber91}%
  \BibitemOpen
  \bibfield  {author} {\bibinfo {author} {\bibfnamefont {L.}~\bibnamefont
  {Weber}}\ and\ \bibinfo {author} {\bibfnamefont {E.}~\bibnamefont {Gmelin}},\
  }\bibfield  {title} {\enquote {\bibinfo {title} {{Transport properties of
  silicon}},}\ }\href {\doibase 10.1007/BF00323873} {\bibfield  {journal}
  {\bibinfo  {journal} {Appl. Phys. A}\ }\textbf {\bibinfo {volume} {53}},\
  \bibinfo {pages} {136--140} (\bibinfo {year} {1991})}\BibitemShut {NoStop}%
\bibitem [{\citenamefont {Wooten}(1972)}]{wooten}%
  \BibitemOpen
  \bibfield  {author} {\bibinfo {author} {\bibfnamefont {F.}~\bibnamefont
  {Wooten}},\ }\href@noop {} {\emph {\bibinfo {title} {Optical Properties of
  Solids}}}\ (\bibinfo  {publisher} {Academic Press},\ \bibinfo {address} {New
  York},\ \bibinfo {year} {1972})\ pp.\ \bibinfo {pages} {244--250}\BibitemShut
  {NoStop}%
\end{thebibliography}
%

%


\newpage
\vspace*{-2.1cm}
\hspace*{-2.5cm}
{
  \centering
  \includegraphics[width=1.2\textwidth,page=1]{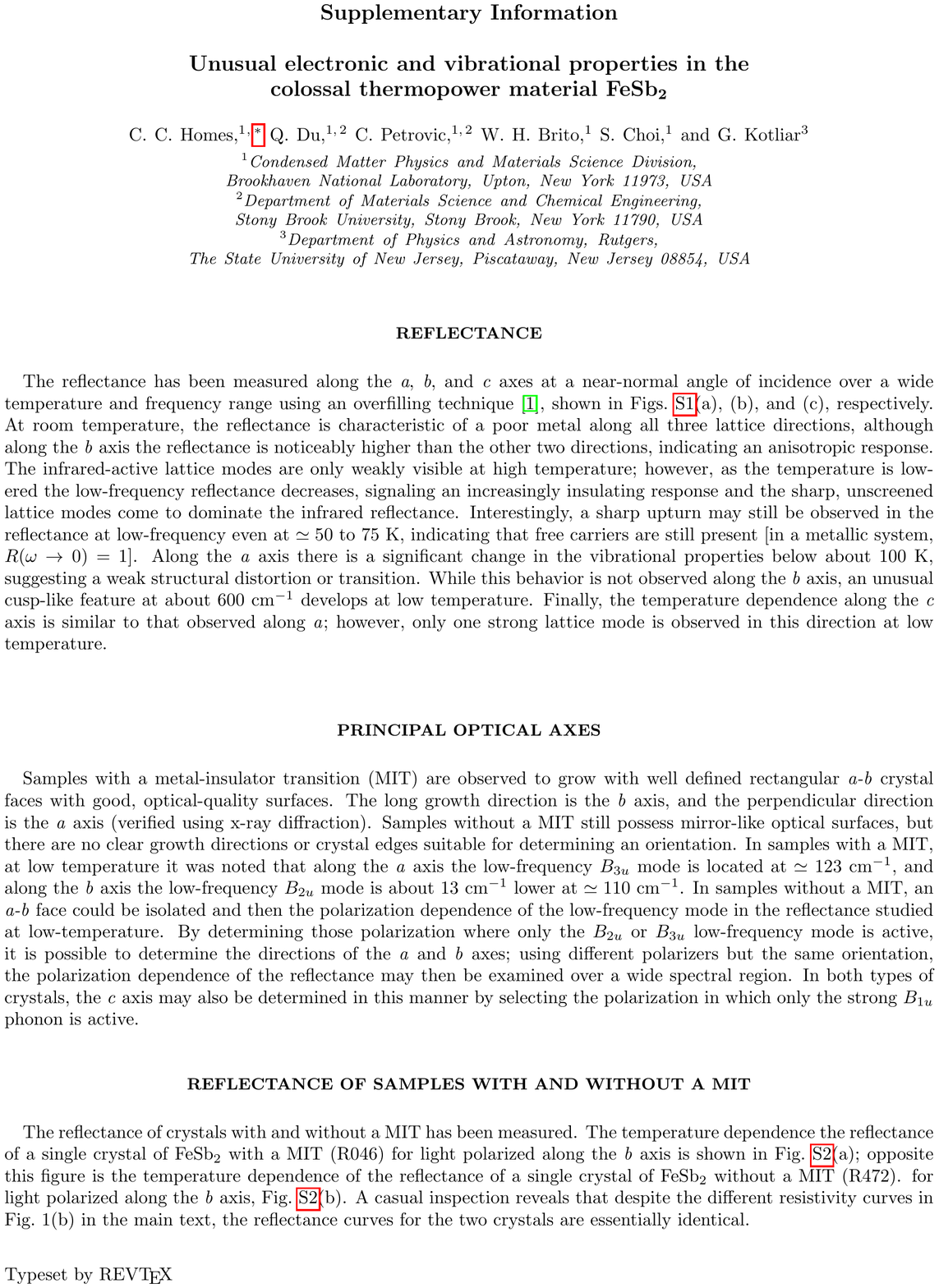} \\
  \ \\
}

\newpage
\vspace*{-2.1cm}
\hspace*{-2.5cm} {
  \centering
  \includegraphics[width=1.2\textwidth,page=2]{supplementary.pdf} \\
  \ \\
}

\newpage
\vspace*{-2.1cm}
\hspace*{-2.5cm} {
  \centering
  \includegraphics[width=1.2\textwidth,page=3]{supplementary.pdf} \\
  \ \\
}

\newpage
\vspace*{-2.1cm}
\hspace*{-2.5cm} {
  \centering
  \includegraphics[width=1.2\textwidth,page=4]{supplementary.pdf} \\
  \ \\
}

\newpage
\vspace*{-2.1cm}
\hspace*{-2.5cm} {
  \centering
  \includegraphics[width=1.2\textwidth,page=5]{supplementary.pdf} \\
  \ \\
}

\newpage
\vspace*{-2.1cm}
\hspace*{-2.5cm} {
  \centering
  \includegraphics[width=1.2\textwidth,page=6]{supplementary.pdf} \\
  \ \\
}

\newpage
\vspace*{-2.1cm}
\hspace*{-2.5cm} {
  \centering
  \includegraphics[width=1.2\textwidth,page=7]{supplementary.pdf} \\
  \ \\
}

\newpage
\vspace*{-2.1cm}
\hspace*{-2.5cm} {
  \centering
  \includegraphics[width=1.2\textwidth,page=8]{supplementary.pdf} \\
  \ \\
}

\newpage
\vspace*{-2.1cm}
\hspace*{-2.5cm} {
  \centering
  \includegraphics[width=1.2\textwidth,page=9]{supplementary.pdf} \\
  \ \\
}

\end{document}